\documentclass[aps,prb,twocolumn,superscriptaddress,showpacs]{revtex4}
\usepackage{amssymb,graphicx,xspace,color}
\newcommand{\tn}{$T_N$\xspace}
\newcommand{\ts}{$T_S$\xspace}
\newcommand{\lao}{LaFeAsO\xspace}
\newcommand{\mn}{$\rm Mn_3Si$\xspace}

\begin{document}
\title{Spin density wave order and fluctuations in $\rm\bf Mn_3Si$: a transport study}


\author{Frank Steckel}
\email[]{f.steckel@ifw-dresden.de}
\author{Steven Rodan}
\author{Regina Hermann}
\author{Christian G. F. Blum}
\affiliation{Leibniz-Institute for Solid State and Materials Research, IFW-Dresden, 01171 Dresden, Germany}
\author{Sabine Wurmehl}
\affiliation{Institut f\"ur Festk\"orperphysik, TU Dresden, 01069 Dresden, Germany}
\affiliation{Leibniz-Institute for Solid State and Materials Research, IFW-Dresden, 01171 Dresden, Germany}
\author{Bernd B\"uchner}
\author{Christian Hess}
\email[]{c.hess@ifw-dresden.de}
\affiliation{Leibniz-Institute for Solid State and Materials Research, IFW-Dresden, 01171 Dresden, Germany}
\affiliation{Center for Transport and Devices, Technische Universit\"at Dresden, 01069 Dresden, Germany}
\date{\today}
\begin{abstract}
We present a comprehensive transport investigation of the itinerant antiferromagnet Mn$_\mathrm{3}$Si which undergoes a spin density wave (SDW) order below $T_N\sim 21.3$~K.
The electrical resistivity, the Hall-, Seebeck and Nernst effects exhibit pronounced anomalies at the SDW transition, while the heat conductivity is phonon dominated and therefore is insensitive to the intrinsic electronic ordering in this compound. At temperatures higher than \tn our data provide strong evidence for a large fluctuation regime which extends up to $\sim200$~K in the resistivity, the Seebeck effect and the Nernst effect. From the comparison of our results with other prototype SDW materials, viz. \lao and Chromium, we conclude that many of the observed features are of generic character.
\end{abstract}

\maketitle

\section{Introduction}
Intrinsic electronic ordering phenomena have in recent years been a focus of condensed matter research in the context of unconventional quantum phenomena, e.g. unconventional superconductivity. For example, in the cuprate high-temperature superconductors it is well established that electronic ordering states, which give rise to inhomogeneous charge and spin distributions, exist and seemingly compete with the superconducting state.\cite{Tranquada95,Tranquada2004,Fink2011,Laliberte2012} Another important material class is that of the more recently discovered iron-pnictide superconductors,\cite{Kamihara2008} where superconductivity emerges upon the suppression of a spin density wave (SDW) state and which suggests that the magnetic and superconducting ground states compete for the electrons near the Fermi level.\cite{Klauss2008,Luetkens2009, Fernandes2010a, PhysRevB.81.140501} Electronic ordering states such as SDW or charge density waves are intimately connected with reconstructions of the Fermi surface topology with respect to the non-ordered states, which cause anomalous behavior of many physical properties at the phase transition. The transport properties are of fundamental importance as the electrons at the Fermi level are directly probed. This concerns in addition to the well known quantities resistivity, Hall and Seebeck effects, also the Nernst effect, which came into focus recently because of its sensitivity to subtle Fermi surface changes and fluctuations.\cite{Bel2003, Bel2004, Chang2010, ISI:000265193600036, Daou2010, Hess2010, kondrat.nernst, PhysRevB.81.045102}

In this paper we take the impact of SDW ordering on the transport coefficients under scrutiny by investigating the transport properties of the itinerant antiferromagnet \mn which undergoes a SDW transition at about 25~K.\cite{tomiyoshi.helicalspin, tomiyoshi.magn.exci.mn3si, tomiyoshi.highsdw, tomiyoshi.magn.ecx.mn3si,pfleid.magn.mn3si} We study in particular the electrical resistivity, thermal conductivity, Hall, Seebeck and Nernst effects in the temperature range from $10$~K up to $300$~K. Clear anomalies are observed at the SDW transition which confirm it to be at $\sim 22$~K and give strong evidence for a large fluctuation regime which extends up to $\sim200$~K in the resistivity, as well as the Seebeck and Nernst effects. We compare our results with other prototype SDW materials, viz. the iron arsenide \lao and the classical SDW prototype Chromium.

\mn is an intermetallic compound with a lattice constant of $a=5.722\;\mathrm{\mathring{A}}$.\cite{aronson} It belongs to the broad family of L2$_\mathrm{1}$ Heusler compounds. In the typical Heusler notation the compound is written as $\mathrm{Mn_2^{II}Mn^{I}Si}$ with two different crystallographic manganese sites. The structure is described by four fcc-lattices with the following positions: $\mathrm{Mn^{I}}$ at (0,0,0), $\mathrm{Mn^{II}}$ at ($\frac{1}{4}$,$\frac{1}{4}$,$\frac{1}{4}$) and ($\frac{3}{4}$,$\frac{3}{4}$,$\frac{3}{4}$) and Si at ($\frac{1}{2}$,$\frac{1}{2}$,$\frac{1}{2}$) in units of the lattice constant $a$.  Through the different surroundings of the Mn-atoms they have different magnetic moments $\mu_\mathrm{Mn^{I}}=1.72\;\mathrm{\mu_B}$ and ${\mu_\mathrm{Mn^{II}}=0.19\;\mathrm{\mu_B}}$ found by neutron diffraction.\cite{tomiyoshi.helicalspin} Additional to the asymmetry of the magnetic moments neutron diffraction experiments revealed an incommensurable SDW with the wave vector ${\vec{q}=4.25\cdot 2\pi/a\cdot 
(1,1,1)}$.\cite{yamaguchi.neutrons.mn3si} Thus \mn is an itinerant antiferromagnet with an incommensurate spin spiral structure.\cite{tomiyoshi.helicalspin, tomiyoshi.magn.exci.mn3si, tomiyoshi.highsdw, tomiyoshi.magn.ecx.mn3si}. Aside from these experimental results, theoretical work suggests the weak magnetic moment of $\mathrm{Mn^{II}}$ to be induced by the $\mathrm{Mn^{I}}$ moment\cite{hortamani.origin.spin.spiral}, consistent with the K\"ubler rule.\cite{Wurmehl2006,Kubler1983} Further theoretical work predicted two nesting vectors of which one corresponds to the experimentally found one.\cite{mohn.supanetz, vlasov} Early publications concerning \mn suggested it to be a possible candidate for half-metallic antiferromagnetism\cite{leuken.hm.antifm, pfleid.magn.mn3si, Jeong2012888, doerr}, which would represent a new paradigm of itinerant magnetism. However, experimental evidence for such a ground state remains elusive.

\section{Experimental}
The \mn single crystalline samples were grown\cite{Hermann2012} using a two-phase radio frequency floating-zone method, described in detail elsewhere.\cite{Hermann2005e1533} The orientation of the crystals was determined using the X-ray Laue back scattering method. The magnetic susceptibility was measured in a superconducting quantum interference device type magnetometer (SQUID, Quantum Design) to further specify the sample properties.\cite{Hermann2012}
The specific heat measurements were performed with a Physical Property Measurement System (PPMS, Quantum Design). All measurements shown here are performed on the same \mn sample with a cuboid shape of the size $\rm 0.5~x~0.5~x~2.25~mm^3$. Electric and thermal currents were forced along the long axis of the crystal which was cut to be the $[110]$-axis. Except for the specific heat all data were taken in a homemade device. The resistivity and Hall measurements were performed as a function of temperature using a standard four-probe technique. During the Hall effect measurements the transverse resistivity $\rho_{xy}$ was linear up to 15~T. All electrically conducting contacts were made using a silver epoxy. For the heat conductivity, the Seebeck and the Nernst effect measurement we used a chip resistor as heater in a steady-state method and a Au-Chromel differential thermocouple to measure the temperature gradient $\nabla T$ along the sample.\cite{PhysRevB.68.184517} We measured the Seebeck effect (also called 
thermopower) at the same time as the heat conductivity by attaching two electrical contacts to the sample along the temperature gradient. The Nernst effect 
was measured in magnetic fields up to $14$~T, with the electrical contacts perpendicular to the thermal gradient. The magnetic field was applied perpendicular to these two directions, and the Nernst signal was linear in field.

\section{Results}
\subsection{Heat capacity}
\begin{figure}
\includegraphics[clip,width=1\columnwidth]{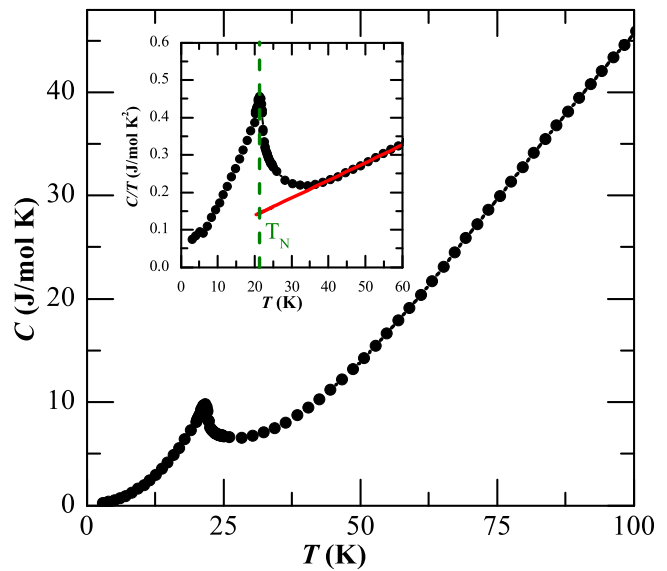}
\caption{(Color online) Heat capacity of \mn (dots). Inset: Heat capacity divided by temperature (dots) and extrapolation above \tn (line) as a guide to the eye}.
\label{fig:figure1}
\end{figure}
The heat capacity is shown in FIG.~\ref{fig:figure1}. Except for the pronounced anomaly with a maximum at $T=21.3$~K the heat capacity is monotonically rising with increasing temperature. It resembles the behavior seen in earlier measurements on polycrystals.\cite{pfleid.magn.mn3si} 
In order to determine the magnetic ordering temperature we refrain from applying the entropy conserving construction. Instead of a smeared out and rather broad peak structure which one would expect for a canonical second order type transition, the anomaly possesses a characteristic $\lambda$-shape, which points towards strong fluctuations (inset of FIG.~\ref{fig:figure1}). In this case we find the transition temperature from the paramagnetic to the SDW region to be exactly at the peak temperature of the anomaly and thus \tn$\simeq 21.3$~K, which agrees rather well with previous reports.\cite{pfleid.magn.mn3si, tomiyoshi.helicalspin} The entropy conserving construction would yield a somewhat higher \tn around $\sim 24$~K.

\subsection{Resistivity}
\begin{figure}
\includegraphics[clip,width=1\columnwidth]{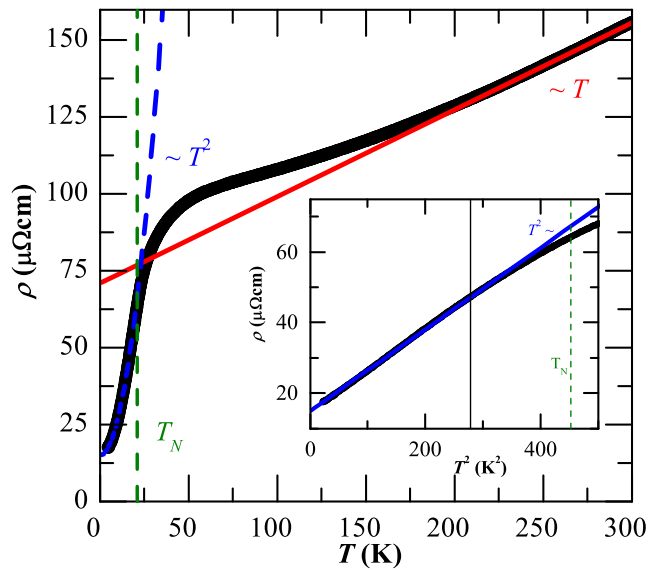}
\caption{(Color online) Resistivity of \mn (solid line), the extrapolated linear high temperature behavior (solid straight line) and the quadratic low temperature behavior (dashed line). The transition temperature (dashed vertical line) is indicated. Inset: low temperature resistivity versus $T^2$ to demonstrate quadratic resistivity behavior below \tn.} 
\label{fig:figure2}
\end{figure}
\begin{figure}
\includegraphics[clip,width=1\columnwidth]{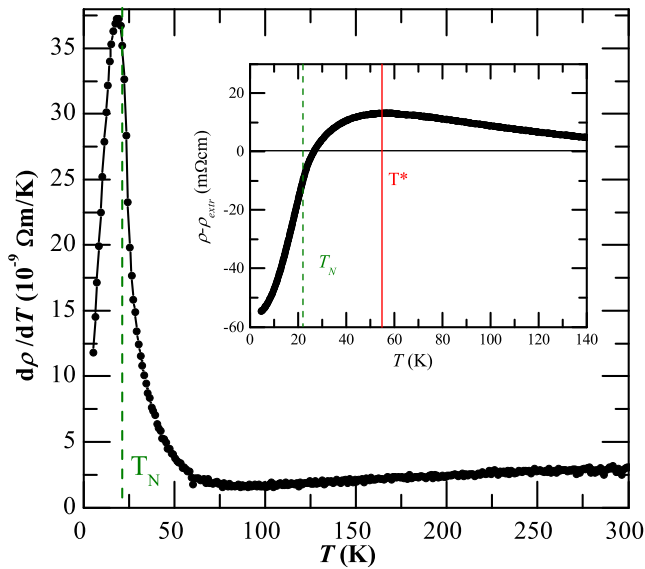}
\caption{(Color online) Derivation of the resistivity of \mn (dots). Inset: resistivity deviation from the extrapolated high temperature behavior $\rho-\rho_\mathrm{extr}$ (dots) solid line marks the temperature $T^*$ with maximum deviation.} 
\label{fig:figure7}
\end{figure}
In FIG.~\ref{fig:figure2} the resistivity is plotted as a function of temperature. The absolute value at room temperature is $\rho(300\;\mathrm{K})=160\;\mu \Omega$cm and the extrapolated residual resistance is $\rho_0=14.88\;\mu \Omega$cm. Our measurement resembles previous results on polycrystalline \mn.\cite{pfleid.magn.mn3si} At temperatures higher than $\sim 200$~K the resistivity approaches a linearly rising behavior with temperature, a typical characteristic of electron-phonon-scattering (expressed in the high temperature limit of the Gr\"uneisen-Bloch formula\cite{ziman} as is indicated by the solid line in FIG.~\ref{fig:figure2}). Apparently the resistivity deviates from this linear behavior towards higher values in a temperature region between 25~K and 200~K, where the maximum deviation from the extrapolated linear $T$-dependence is at around $T^*=55$~K (cf. inset of FIG.~\ref{fig:figure7}). At $T<T^*$ the resistivity starts to decrease stronger and then crosses the extrapolated high temperature linear $T$-dependence close to \tn. At temperatures below \tn the resistivity decreases further, exhibits an inflection point at $T\sim 20$~K, i.e. nearly exactly at \tn and shows a crossover to a $T^2$-dependence which is complete at $T\lesssim14$~K. 

The deviation of the resistivity from the expected linear high temperature behavior can only arise if the number of charge carriers or their relaxation time and thus their scattering probability changes. Due to the very small change of the Hall coefficient (see next section) we attribute the deviation to magnetic fluctuations which are indicated by the enhanced scattering of the charge carriers. Below $T^*$ the additional scattering reduces and apparently vanishes at \tn. This suggests that the enhanced scattering is intimately connected to the SDW transition, where it seems natural to assign the enhanced scattering to nesting related processes, and on the other hand the reduction of scattering to the incipient magnetic order below $T^*=55$~K. Note that the width of the transition at \tn as observed in the $c_p$ data corresponds roughly to this temperature regime. 

The strong decrease  of $\rho$ below \tn indicates a drastic reduction of the carrier scattering which even overcompensates the reduced carrier density due the opening of a SDW gap. Note that this gap opening seems to be completed at the inflection point at $\sim 20$~K, where the decrease of the resistivity becomes weaker (cf. FIG.~\ref{fig:figure7}). This marks the completion of the transition at \tn. A natural explanation for reduced scattering below \tn is a reduction of phase space upon the SDW transition. The $T^2$-dependence of the resistivity at lowest temperature is most reasonably explained by scattering on magnetic one-particle fluctuations (magnons) in the antiferromagnetic phase.\cite{Bombor2013,JPSJ.43.1497, PSSB:PSSB19670210236} 

It is interesting to compare these findings with resistivity data of other prototype SDW materials. The material \lao may be viewed as a representative case of the iron arsenide parent compounds exhibiting SDW order. Its magnetic transition occurs at \tn$=137$~K which is preceded by a structural transition from tetragonal to orthorhombic at $T_S=160$~K.\cite{0295-5075-87-1-17005,mcguire.lafeaso}. The temperature dependence of the resistivity of this compound\cite{hess2012} is remarkably similar to that of \mn. This concerns almost all the  qualitative observations except a low-temperature upturn which is present in the resistivity of \lao, i.e. the linear high temperature behavior, the enhanced scattering above \tn and the strong reduction below, including the inflection point.

The SDW order in \mn has often been compared with the elemental SDW material Cr.\cite{tomiyoshi.helicalspin,pfleid.magn.mn3si} Surprisingly the temperature behavior of the resistivity of Cr is very different because it exhibits a small hump below \tn.\cite{chrom.moore, fawcett.review.chrom}

\subsection{Hall effect}
\begin{figure}
\includegraphics[clip,width=1\columnwidth]{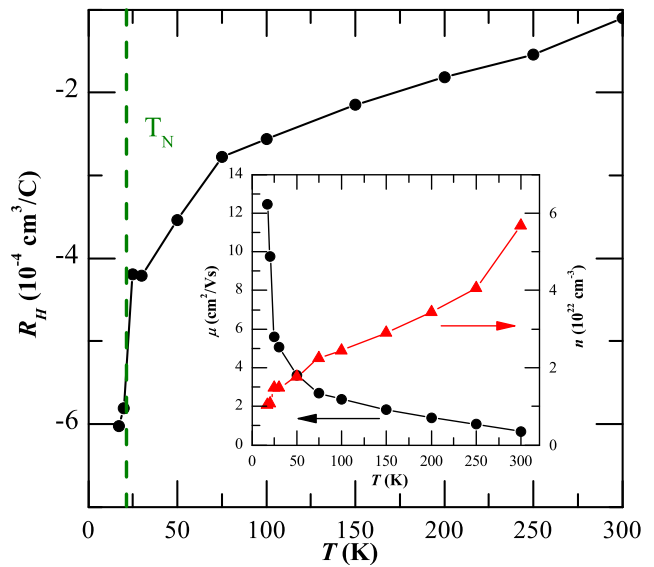}
\caption{(Color online) Hall effect of \mn. Inset: calculated charge carrier mobility (dots) and charge carrier density (triangles) in an assumed one carrier type case.} 
\label{fig:figure3}
\end{figure}
The temperature dependence of the Hall coefficient $R_H$ is shown in FIG.~\ref{fig:figure3}. $R_H$ is negative over the complete temperature range, which in a one-band model corresponds to electrons as charge carriers. Note that the one band picture is a simplified approach because \mn is known to be a multiband metal from band structure calculations.\cite{fujii.theory.hm} 

At 300~K the Hall coefficient is ${R_H=-1\cdot 10^{-4}\;\mathrm{cm}^{-3}\mathrm{/C}}$ and increases almost linearly to more negative values upon decreasing the temperature down to around 75~K where ${R_H\approx -2.8\cdot 10^{-4}\;\mathrm{cm}^{-3}\mathrm{/C}}$. Such a weak temperature dependence of $R_H$ is characteristic for multiband materials. Below 75~K and again below \tn the slope of $R_H(T)$ changes towards larger positive values which is connected to large negative values of $R_H$ ($R_H\approx -6\cdot 10^{-4}\;\mathrm{cm}^{-3}\mathrm{/C}$ at 17~K).

The drop of $R_H$ below \tn can clearly be attributed to the SDW order. The qualitative origin of this drop can straightforwardly be connected with the opening of a SDW gap. The slope change below 75~K corresponds roughly to the region of incipient magnetic order which above was identified  by the reduction of the additional scattering in the resistivity at $T\lesssim T^*$, cf. FIG.~\ref{fig:figure7}.

We find that the mobility $\mu=R_H/\rho$ strongly increases with decreasing temperature, and again assuming a one-band picture one can extract the carrier density $n$. Both quantities are shown in the inset of FIG.~\ref{fig:figure3}.
The carrier density at room temperature is found to be $n=5.7\cdot 10^{22}\;\mathrm{cm}^{-3}$. This value is somewhat lower than that of elemental metals (K, Na)\cite{kittel} and thus, connected with the relatively large value of the resistivity at room temperature, \mn may be qualified as a poor metal.

At this point it is again instructive to compare these findings with the Hall data of \lao.\cite{kondrat.nernst} There, the Hall coefficient is in a similar way only weakly temperature dependent above the transition temperature. Below the closely connected \tn and \ts, the absolute value of the Hall coefficient increases by roughly one order of magnitude. Thus, except for the absolute values we have practically the same behavior in both \mn and \lao.

Early Hall effect measurements\cite{Vries} on Cr seem to yield a similar anomaly at the SDW transition also in this compound, indicative of a significant change of the carrier density at the transition.

\subsection{Thermal conductivity} 
\begin{figure}
\includegraphics[clip,width=1\columnwidth]{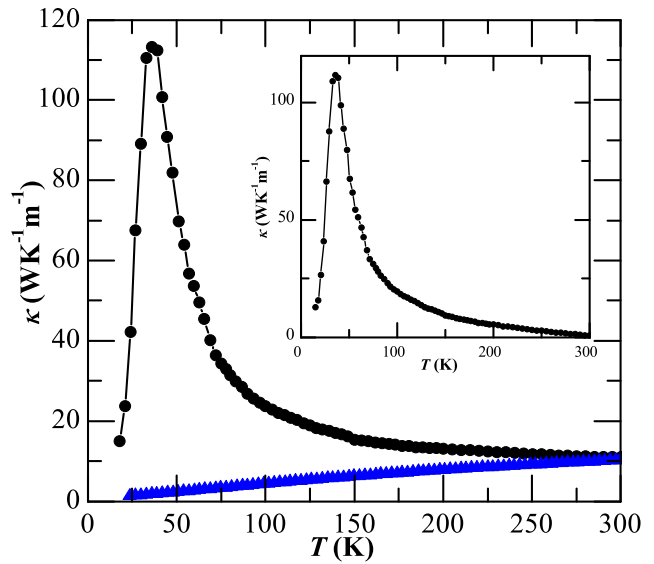}
\caption{(Color online) Thermal conductivity of \mn (dots), electronic contribution calculated by the Wiedemann-Franz-law (triangles). Inset: thermal conductivity of \mn subtracted by the electronic contribution (dots).}
\label{fig:figure4}
\end{figure}
The thermal conductivity plotted in FIG.~\ref{fig:figure4} is nearly constant for temperatures above 150~K with a value of about $\kappa = 10\;\mathrm{WK^{-1}m^{-1}}$. Below this temperature $\kappa$ rises to a maximum of $\kappa_\mathrm{max} = 114\;\mathrm{WK^{-1}m^{-1}}$ at 37~K and goes down rapidly for $T\rightarrow 0$. To understand the contributions to the thermal conductivity we estimated at first the electronic contribution by the Wiedemann-Franz-law:
\begin{equation}
\kappa=L\sigma T
\end{equation}
with $L= \frac{\pi^2}{3}\left(\frac{k_B}{e}\right)^2=2.45\cdot 10^{-8}\;\frac{\mathrm{W\Omega}}{\mathrm{K^2}}$, the theoretical result of the Drude-Sommerfeld-theory. This yields (see triangles in FIG.~\ref{fig:figure4}) a relatively small contribution at low temperatures ($T\lesssim150$~K). However at higher temperatures it increasingly approaches the measured $\kappa$, which is a sign that the electronic contribution is overestimated in this temperature regime. The inset of FIG.~\ref{fig:figure4} shows the measured data with the electronic part subtracted. As expected, the resulting heat conductivity corresponds well to that of a typical phononic heat conductor. Thereby we assume that \mn does not have a contribution of magnons. We did not observe a significant anomaly below \tn. Such an anomaly could in principle arise from magnetoelastic coupling. A further analysis of this rather featureless phonon dominated heat conductivity of \mn and a comparison with that of other SDW materials is therefore concluded to not provide further insights.

\subsection{Seebeck effect}
\begin{figure}
\includegraphics[clip,width=1\columnwidth]{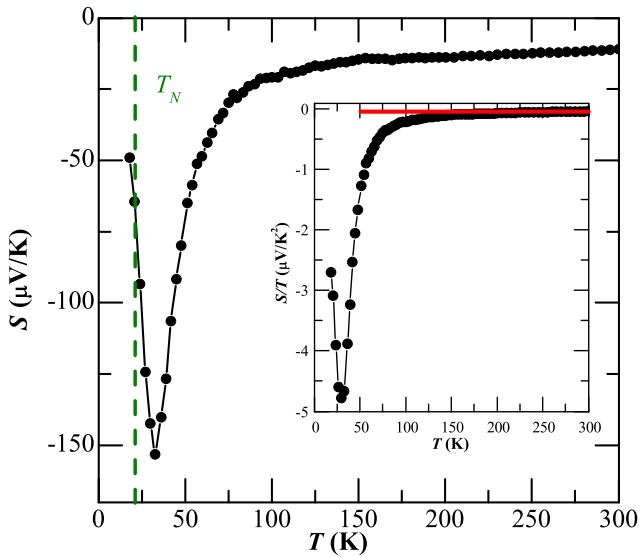}
\caption{(Color online) Thermopower of \mn (dots). \tn is marked with a dashed line. Inset: Thermopower divided by temperature (dots) and the constant value of $S/T=-0.04\;\mathrm{\mu V /K^2}$ (solid line).}
\label{fig:figure5}
\end{figure}
The Seebeck coefficient (see FIG.~\ref{fig:figure5}) is negative over the whole measured temperature range, consistent with the negative sign of $R_H$. The Seebeck coefficient of \mn has a value of $S=-20\;\mathrm{\mu V /K}$ at 300~K and it continuously falls to more negative values with decreasing temperature towards a pronounced anomaly with $S=-160\;\mathrm{\mu V /K}$ at 33~K approaching $S=0\;\mathrm{\mu V /K}$ for $T\rightarrow0$. The transport equations yield the following expression for the thermopower:\cite{ziman}
\begin{equation}
S=\frac{\pi^2}{3}\frac{k_B^2T}{q}\left[\frac{\partial\, ln\, \sigma (E)}{\partial E}\right]_{E=E_F}
\label{eqn:S}
\end{equation}
where $q$ denotes the charge of the carriers, $\sigma(E)$ stands for the electrical conductivity in dependence of the energy.\cite{ziman} Since in Eq.~(\ref{eqn:S}) the Seebeck coefficient is depending linearly on the temperature, it is worthwhile to analyze $S/T$ (see inset of FIG.~\ref{fig:figure5}). The strong temperature dependence of this quantity apparently has to be ascribed to the quantity $\frac{\partial\, ln\, \sigma (E)}{\partial E}$ which in the case of a momentum independent mean free path $l_e$ may be broken down to:\cite{ziman} 
\begin{equation}
\frac{\partial\, ln\, \sigma (E)}{\partial E}=\frac{\partial\, ln\, l_e}{\partial E}+\frac{\partial\, ln\, A_{FS}}{\partial E}
\label{eqn:S2}
\end{equation}
Here, $A_{FS}$ denotes the Fermi surface area.

Eq.~(\ref{eqn:S2}) suggests that the temperature dependence of $S/T$ can be understood as stemming from separate contributions which are associated with the energy dependence of scattering processes and that of Fermi surface topology changes. If one suspects the changes in the Fermi surface topology associated with the SDW phase transition to occur in a relatively narrow temperature range, then the second term of Eq.~(\ref{eqn:S2}) contributes to the temperature dependence of $S/T$ only in the vicinity of the phase transition. All other temperature dependence is then captured by the first term. At $T\rightarrow\infty$ and at $T\rightarrow0$ one expects the Fermi surface topology to be robust and fluctuations (which presumably contribute to the first term) to be negligible. Thus $S/T$ is expected to approach a constant value in both regimes. For the high-temperature limit this is clearly observed in the data, whereas the low-temperature limit is obviously not reached in the present data. While the pondering of these limits yields a clear-cut physical picture, it is impossible to disentangle contributions of the two terms in Eq.~(\ref{eqn:S2}) in the vicinity of the SDW transition where a strong temperature dependence is observed. For temperatures above \tn, $S/T$ strongly deviates at $T\lesssim200$~K from the high-temperature limit which has to be ascribed to both changes of electron scattering and the Fermi surface topology fluctuations. Scattering processes and fluctuations freeze out at \tn and therefore $S/T$ is expected to rapidly approach the low-temperature limit. This is reflected in the strong changes of $S/T$ and the observed minimum. Note that the minimum is at a significantly larger temperature than \tn which means that the Seebeck coefficient responses already to a finite correlation length at temperatures well above the ordered regime.
 
It is well known that in addition to these purely electronic effects the electron-phonon drag might play some role in the Seebeck effect. The drag effect becomes observable in a temperature range where the heat conductivity and thus the phononic mean free path is high\cite{ziman}. Interestingly, the phonon heat conductivity (see FIG.~\ref{fig:figure4}) peaks at the same temperature as the Seebeck coefficient. It remains unclear whether this is just coincidental, or if this indicates a significant importance of the electron-phonon drag.

The comparison with Seebeck coefficient data\cite{springerlink:10.1140/epjb/e2009-00267-3, kondrat.nernst,chrom.moore} for \lao and Cr shows that the observed characteristics, namely a fluctuation regime at $T>T_N$ and a sharp change of $S$ at \tn, are apparently generic features at SDW transitions. Interestingly, the contributions in the fluctuation regime and at $T\lesssim T_N$ have the same sign in both \lao and Cr, whereas their sign is opposite in \mn. We attribute these differences to details of the band structure.

\subsection{Nernst effect}
\begin{figure}
\includegraphics[clip,width=1\columnwidth]{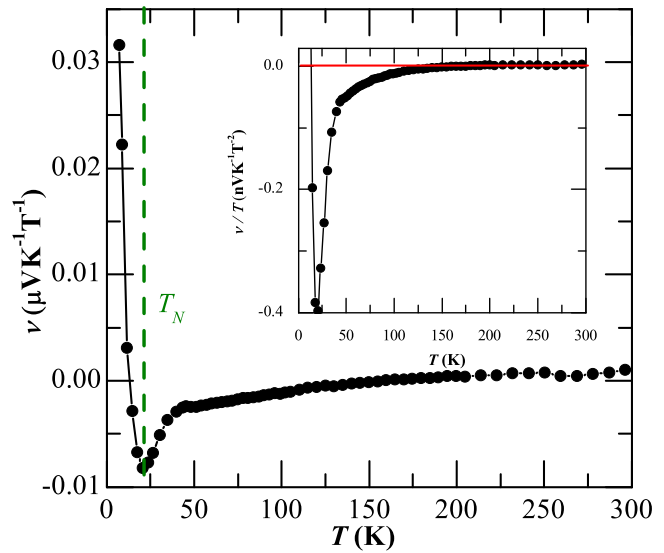}
\caption{(Color online) Nernst coefficient of \mn (dots). Inset: Nernst coefficient divided by temperature.}
\label{fig:figure6}
\end{figure}

For measuring the Nernst effect a temperature gradient is applied along the $x$-direction of the sample in the presence of a magnetic field $B$ along the $z$-axis. The Nernst signal $N$ is a voltage drop as the signature of an electric field $E_y$ along the $y$-direction of the sample.\cite{wang.nernst,behnia.review.nernst} In non-superconducting metals the Nernst signal is expected to be linear in magnetic field. Thus one defines the Nernst coefficient as 
\begin{equation}
 \nu_N=\frac{N}{B}=\frac{E_y}{|-\partial_x\,T|\; B}
\end{equation}
We are using the new sign convention after which a superconducting vortex would give a positive contribution to $\nu_N$.\cite{behnia.review.nernst}
In compounds in which the phononic heat conductivity is far higher than the electronic contribution to the heat conductivity, as it is the case in \mn at low temperature, one can 
write for the Nernst coefficient\cite{PhysRevB.64.224519} 
\begin{equation}
\nu_N =\left[\frac{\alpha_{xy}}{\sigma_{xx}}-S_{xx}\tan \theta\right]\frac{1}{B}
\label{eqn:nu}
\end{equation}
$\alpha_{xy}$ denotes the non-diagonal Peltier coefficient and $\tan \theta = \frac{\sigma_{xy}}{\sigma_{xx}}$ the Hall angle. In a simple one-band metal the two terms on the right hand side are expected to cancel each other out exactly, which is often called the Sondheimer cancellation. It can be shown that in multiband metals and in superconductors in the mixed state Sondheimer's rule is violated.\cite{wang.nernst, behnia.review.nernst} The Nernst coefficient may therefore be considered as a measure to what extend a metal deviates from a simple one-band metal.

An alternative expression for the Nernst coefficient is given by\cite{oganesyan2004,behnia.review.nernst}
\begin{equation}
 \nu_N = -\frac{\pi}{3}\frac{k_B^2T}{eB} \left[ \frac{\partial \tan \theta}{\partial E}\right]_{E=E_F}
\label{eqn:nernst}
\end{equation}
In this formulation Sondheimer's cancellation corresponds to the exact vanishing of the energy dependence of the Hall angle.\cite{behnia.review.nernst} Since the prefactor in Eq.~(\ref{eqn:nernst}) is linear in temperature, and the Hall angle depends on both the carrier scattering rate and their effective mass, one may  qualitatively analyze the temperature dependence of the Nernst coefficient in a similar way as that of the Seebeck coefficient, as will be discussed further below.

In the case of \mn the Nernst coefficient at room temperature is very small $\sim 1\;\mathrm{nVK^{-1}T^{-1}}$, and decreases roughly linearly to zero at about 150~K where $\nu_N(T)$ changes its slope to a somewhat larger value. At 44~K a kink appears and the Nernst coefficient decreases even stronger with decreasing temperature towards a minimum value of  $-8.2\;\mathrm{nVK^{-1}T^{-1}}$ at $\sim 20$~K. This is almost exactly the temperature of the inflection point in the resistivity and that of the maximum of the specific heat anomaly, i.e. \tn. At lower temperatures, and thus deep in the magnetic regime, the Nernst coefficient increases strongly towards $\nu_N=3.16\cdot 10^{-2}\;\mathrm{\mu VK^{-1}T^{-1}}$ at the lowest measured temperature of 7~K.

$\nu_N/T$, which is plotted in the inset of FIG.~\ref{fig:figure6} is very small and temperature independent at $T\geq150$~K, which shows that \mn in this temperature regime behaves as an ordinary metal in line with the linear resistivity at $T>200$~K. The strong temperature dependence at lower temperature according to Eq.~(\ref{eqn:nernst}) stems from the energy dependence of the Hall angle and implies strong changes in the scattering time and the effective mass. It is clear that both quantities experience strong variations in the vicinity of the SDW phase transition, therefore we cannot distinguish between these two contributions in the Nernst coefficient. In a similar way as with the Seebeck coefficient we can understand the temperature dependence of the Nernst coefficient in terms of a fluctuation regime in the range \tn$<T<150$~K and a regime with a qualitatively different behavior at lower temperatures. Note that the latter sets in at \tn in contrast to the minimum of the Seebeck effect.

We point out that the temperature dependence of the Nernst and Seebeck coefficients is qualitatively very similar. Since phonons are not influenced by the magnetic field, this similarity implies that the electron-phonon drag plays only a minor role in the Seebeck coefficient. However, we note that despite the similarities at the high and low-temperature regimes in the intermediary temperature regime
\tn$<T<50$~K the temeprature dependencies of the two effects is remarkably different. This concerns mostly the temperature of the minimum  in the vicinity of \tn, which might indicate a different sensitivity to the correlation length of the incipient SDW order.

The origin of the remarkable enhancement of the Nernst coefficient in the regime $20~{\rm K}\leq T\leq 44$~K is unclear. One may speculate, however, that it corresponds to enhanced fluctuations in the direct vicinity of the phase transition. Interestingly, it roughly coincides with the width of the specific heat anomaly and the temperature regime between \tn and $T^*$ of the resistivity. This illustrates that the measurement of the Nernst effect is a remarkable and powerful complementary method to more conventional transport coefficients which underpins that this quantity is extremely sensitive to fluctuations of the Fermi surface topology.

We again compare these findings with results for \lao,\cite{kondrat.nernst} where it is observed that the Nernst coefficient is nearly constant and zero at $T\gg T_N$. Upon approaching the SDW transition from above, corresponding fluctuations have also been reported to lead to an enhanced Nernst response, which increases even further below the SDW transition. Note that in \lao the fluctuation-enhanced and the SDW-enhanced Nernst coefficient are of the same sign, whereas the respective signs are opposite in \mn. 

\section{Conclusion}
In conclusion, we investigated a comprehensive set of transport coefficients on a single crystalline sample of the itinerant antiferromagnet \mn. All transport coefficients except the thermal conductivity are sensitive to the SDW transition in this material and exhibit strong anomalies around the ordering temperature $T_N\sim 21.3$~K. 
These anomalies qualitatively arise from both strongly temperature-dependent changes of the relaxation time and the Fermi surface topology changes in relation to the SDW transition. 
Such transport investigations are therefore an important and powerful tool for disentangling the nontrivial nature of the magnetism of itinerant electron systems. This is further demonstrated by the apparent generic nature of many of the observed characteristics related to the phase transition, which are deduced from comparison with similar studies on other prototype SDW compounds.
We point out that the rarely studied Nernst effect apparently provides a rather rich spectrum of information which underpins the potential of this quantity for experiments in solid state physics.

An interesting finding which is evident in \mn from the resistivity, Seebeck coefficient and Nernst coefficient data is a large fluctuation regime which extends up to about 200~K. Fluctuations which evolve already at temperatures almost one order of magnitude higher than the actual ordering temperature appear rather unusual for a three dimensional metal. One might speculate that this large fluctuation regime is the signature of competing orders in the compound. This notion is nourished by the theoretical finding of a second nesting vector in the electronic structure which is calculated to cause an even stronger instability than that related to the actual observed order.\cite{vlasov, mohn.supanetz}

\section*{ACKNOWLEDGMENTS}
This work was supported by the Deutsche Forschungsgemeinschaft through SPP 1538, grant HE 3439/9, and through GRK 1621. S. Wurmehl acknowledges funding by DFG in project WU 595/3-1 (Emmy-Noether program).

\end{document}